\documentstyle[twocolumn,psfig,prb,aps]{revtex}

\begin{document}
\bibliographystyle{plain}

\newcommand{\chemical}[1]{{$\rm #1$}}
\newcommand{\frak}{\sf}
\renewcommand{\vec}{\bf}
\newcommand{\altvec}[1]{{\underline{\rm #1}}}
\newcommand{\mat}{\sf\bf}
\newcommand{\altmat}[1]{{\underline{\underline{\rm #1}}}}
\renewcommand{\imath}{{\rm i}}
\newcommand{\emath}{{\rm e}}
\newcommand{\op}{\cal}
\newcommand{\func}{\rm}
\newcommand{\set}{\frak}
\newcommand{\stackindex}[2]{{\stackrel{\scriptstyle #1}{\scriptstyle #2}}}
\newcommand{\avg}[1]{{\bigl< #1 \bigr>}}
\newcommand{\greenf}[2]{{\bigl<\bigl< #1;#2 \bigr>\bigr>}}
\newcommand{\transpose}[1]{{{#1}^{\rm T}}}
\newcommand{\phase}[1]{$\bigl < #1 \bigr >$}
\newcommand{\sign}{{\rm sign~}}
\newcommand{\trace}{{\rm tr~}}

%
\title{Derivation of symmetry-based pseudo spin models for modulated materials}
\author{Boris Neubert, Michel Pleimling${}^\dagger$, and Rolf Siems}
\address{
Theoretische Physik, 
Universit\"at des Saarlandes, 
Pf.\ 151150, 
D-66041 Saarbr\"ucken,
e-mail: neubert@lusi.uni-sb.de\\
${}^\dagger$Theoretische Physik, 
Rheinisch-Westf\"{a}lische Technische Hochschule, 
Sommerfeldstra{\ss}e,  
D-52074 Aachen}
\date{\today}
\maketitle

\begin{abstract}
A general concept for the derivation of symmetry-based pseudo spin Hamiltonians is 
described. It systematically bridges the gap between the atomistic
basis and various pseudo spin models presented in literature. 
It thus allows the application of the
multitude of analytical and numerical procedures derived for the statistical mechanics
treatment of the latter to the description of structurally modulated crystals and
furnishes a general frame for reviewing these models from a common point of view.
\end{abstract}


\section{Introduction}
%
Various approaches for the description
of systems exhibiting 
commensurately and/or incommensurately structurally modulated phases were presented
in the literature:
Landau-type models\cite{San90b,Rib90},
the model of Chen and Walker\cite{Che91a,Che91b} for 
\chemical{A_2BX_4}-compounds and betaine calcium chloride dihydrate (BCCD),
the model of Hlinka et al.\cite{Hli96} for BCCD, other
microscopic models with continuous local variables like the DIFFOUR (discrete
$\Phi^4$) models\cite{Jan81a} and pseudo spin models like
the ANNNI (axial next nearest neighbor Ising)\cite{Ell61,Sel88}, AANNDI (axial 
antisymmetric nearest neighbor double Ising)\cite{Kur86} and DIS (double Ising 
spin)\cite{Ple94,Ple97a} models.
Ab initio and molecular dynamics (MD)
investigations\cite{Lu90,Etx92a} allow to gain insight into processes and interactions on 
an atomistic level.\par
A comparative analysis of the merits and shortcomings of these 
models is facilitated and systematized  by the development of
a general procedure for the formulation of model Hamiltonians
for uniaxially structurally modulated compounds. Such a procedure is described in this
paper. It is based on the discrete arrangement of atoms and makes use
of the symmetry of the lattice.
It leads to a hierarchy of Hamiltonians corresponding to different approximations, 
that describe, among others, the different microscopic 
models mentioned above. These can thus be
discussed from a common point of view.
The models are obtained by identification of the relevant variables 
and specification of the respective couplings. 
Their Hamiltonians allow both numerical and approximate analytical treatments. 
One advantage of such atomistic symmetry-adapted model Hamiltonians is that they allow
predictions also on complicated phase diagrams containing
long-period structures, on their space group symmetries, and on phase transitions 
which are not accessible by ab initio and MD calculations.
Previously it was considered a weak spot, especially of pseudo spin models, 
that the significance (on an atomistic scale) of 
the model (pseudo spin) variables was not well defined and that predictions
in terms of space group symmetries could not be made\cite{Che91a,Sch93}.
%
\section{Formulation in terms of symmetry-adapted local modes}
%
The basis of the procedure is to start from
symmetry-adapted {\em local} modes (SALMs) as a localized basis set for the expansion 
of atomic displacements in properly chosen crystallographic (sub)cells.
This facilitates the incorporation of both the discreteness of the
lattice and the overall symmetry of the crystal. The {\em relevant}
SALMs can either be determined by first-principles calculations or taken from
the observed displacements at the phase transitions.
Generalized local variables defined in terms of the amplitudes of these SALMs serve as
variables for a microscopic Hamiltonian.
This procedure stresses the atomistic roots of the models, leads to models
conforming to lattice symmetry and facilitates the prediction of spontaneous
polarizations of modulated phases and of the displacements occuring at structural 
transitions.\par 
Let the crystal be composed of $N$ unit cells with $K$ particles each 
(of which $K'$ are non equivalent).
The crystal is described by a $3KN$-dimensional configuration vector 
${\altvec R}= {\altvec R}_0 + \delta{\altvec R}$, whose entries specify the 
average positions
of the atoms. ${\altvec R}_0$ describes a $T$- and $p$-dependent reference structure
which is invariant under all elements
${\op G}=\{{\mat R}|{\vec t}\}$ of the normal phase space 
group $\frak G_0$. The symmetry-breaking displacement $\delta{\altvec R}$ is zero 
in the normal phase.
In the lower symmetry phases (unmodulated or commensurately or incommensurately modulated), 
it can be written as a superposition 
\begin{equation}
\delta{\altvec R}= \sum_{\vec n} \delta{\altvec R}_{\vec n}
\label{equation:Wexpansion}
\end{equation}
of local contributions $\delta{\altvec R}_{\vec n}$ having non-zero entries 
for only those atoms which are associated to cell $\vec n$.\par
$\delta{\altvec R}_{\vec n}$ is expanded in terms of a local basis set.
If the set reflects the crystal symmetries, it is sufficient in most cases 
to retain only a few (e.g.\ one or two) local modes.
The smallest unit in the crystal with respect to symmetry 
is the asymmetric unit ${\frak A}^1_{\vec 0}$ of the space group $\frak G_0$.
Symmetry imposes no restrictions on the positions of the $K'$ atoms associated
to this polyhedral subcell (nor need they to be known except for a quantitative atomistic
determination of model parameters).  Further subcells ${\frak A}^i_{\vec n}$ defined by
\begin{displaymath}
{\frak A}^i_{\vec 0}= {\op G}^i {\frak A}^1_{\vec 0} \qquad\mbox{and}\qquad {\frak A}^i_{\vec n}=\{{\mat E}|{\vec n}\} {\frak A}^i_{\vec 0}
\end{displaymath}
cover the whole crystal. The
$B=[{\frak G_0}:{\frak T}]$ space group operations ${\op G}^1,\ldots,{\op G}^B$ are coset 
representatives in a coset decomposition of $\frak G_0$ with respect to its 
subgroup $\frak T$ of primitive translations.  
They have to be chosen such that the subcells 
${\frak A}^1_{\vec n},\ldots,{\frak A}^B_{\vec n}$ form a contiguous new choice of cell 
$\vec n$. 
The basis set of $3K'$ subcell modes 
${\altvec V}^{i\kappa}_{\vec n}, \kappa=1,\ldots,3K'$ for the displacements of the $K'$ 
atoms associated with ${\frak A}^i_{\vec n}$ should satisfy the relation
${\altvec V}^{i'\kappa}_{\vec n'}={\op G}{\altvec V}^{i\kappa}_{\vec n}$, where
${\op G}\in{\frak G_0}$ is the space group operation transforming 
${\frak A}^{i}_{\vec n}$ into ${\frak A}^{i'}_{\vec n'}$. It is thus 
entirely determined by the choice of modes in the asymmetric unit 
${\frak A}^1_{\vec 0}$.\par
A complete -- for $K'B>K$ over-complete (see below) -- set of $3K'BN$ local 
modes can be constructed as linear combinations 
\begin{equation}
\label{equation:WSV}
{\altvec W}^{I\kappa}_{\vec n}=\sum_{i=1}^B S^{Ii} {\altvec V}^{i\kappa}_{\vec n}
\end{equation}
of the modes in subcells $i=1,\ldots,B$ with a $B \times B$ matrix $\altmat S$. 
The subset $\{{\altvec W}^{I\kappa}_{\vec n}: \kappa= 1, \ldots, 3K'\}$ is called the
$I$th set of SALMs for cell $\vec n$. 
The transformation matrix
$\altmat S$ should be chosen in accordance with the irreducible representations (IRs)
of the point group $\widehat{\frak G}_0$ such that for the construction of every
symmetry mode (deformation $\delta\altvec R$ transforming
according to an IR of the group $\frak G({\vec q})$ of the wave vector) only a 
minimum of different sets of SALMs is needed.
Local modes ${\altvec W}^{I\kappa}_{\vec n}$ and ${\altvec W}^{I\kappa}_{\vec m}$
describe the {\em same} displacement patterns
in their respective cells, 
i.e.\ ${\altvec W}^{I\kappa}_{{\vec m}}=\{{\mat E}|{\vec m}-{\vec n}\} {\altvec W}^{I\kappa}_{\vec n}$.
All linear combinations
of displacement vectors in one set of SALMs have the same transformation behavior,
which is entirely determined by $\altmat S$, since,
apart from a primitive translation, an 
arbitrary space group operation only permutes the subcell indices. 
Only those sets of SALMs can contribute to the deformation $\delta\altvec R$ that
have the symmetry properties imposed by the IR according to which $\delta\altvec R$ 
transforms.\par
Atoms on special positions (faces, edges and corners of the
subcells) cause the number $3K'BN$ of local modes to be higher than the number
$3KN$ of independent displacements. Hence, the coefficients of an expansion 
in terms of these SALMs have to satisfy consistency relations.\par
The symmetry-breaking displacement $\delta{\altvec R}_{\vec n}$  in cell $\vec n$
is decomposed 
into contributions $\delta{\altvec R}^I_{\vec n}$ from the $B$ sets of SALMs:
\begin{equation}
\label{equation:deltaRn}
\delta {\altvec R}_{\vec n}= \sum_{I=1}^B \delta {\altvec R}_{\vec n}^I= \sum_{I=1}^B \sum_{\kappa=1}^{3K'} a_{\vec n}^{I\kappa} {\altvec W}_{\vec n}^{I\kappa}.
\end{equation}
Under a change of external parameters, $\delta {\altvec R}_{\vec n}$ will
follow a valley of low local potential (see below) in the high-dimensional 
configuration space corresponding to the modes relevant for the phase transitions.
In the  simplest and frequent case this valley is not very shallow and the contribution 
$\delta{\altvec R}_{\vec n}^I = \sum_{\kappa=1}^{3K'} a_{\vec n}^{I\kappa} {\altvec W}_{\vec n}^{I\kappa}$
from the $I$th set of SALMs 
will lie on a curve (dimension $d_a=1$). In a more general case, 
$\delta{\altvec R}_{\vec n}^I$ will lie on a $d_a\leq3K'$-dimensional 
curved surface $\Sigma$. This generalization would account 
for deviations due to {\em different} (small) external fields and allow the treatment
of additional fluctuations (about the bottom
of the valley) described by the additional $d_a-1$ modes.
The special form of the symmetry-breaking
displacement (in the asymmetric unit ${\frak A}^1_{\vec 0}$)
connected with the transitions is not
determined by space group symmetry. It will, in general,  
be {\em different} even for materials
exhibiting the {\em same} symmetries of the normal and the modulated phases 
(e.g.\ \chemical{A_2BX_4} compounds). 
One introduces local sets of curvilinear coordinates for every set of SALMs
by means of a coordinate transformation 
$a_{\vec n}^{I\kappa}= f^{I\kappa}(Q^{I1}_{\vec n},\ldots,Q^{I,3K'}_{\vec n}),\,
\kappa=1,\ldots,3K'$ leading to
\begin{displaymath}
\delta{\altvec R}_{\vec n}^I(Q^{I1}_{\vec n},\ldots,Q^{I,3K'}_{\vec n})=
\sum_{\kappa=1}^{3K'}f^{I\kappa}(Q^{I1}_{\vec n},\ldots,Q^{I,3K'}_{\vec n}){\altvec W}_{\vec n}^{I\kappa}
\end{displaymath}
such that $\Sigma$ is parametrized by the first $d_a$ 
coordinates (relevant variables per set of SALMs).\par
The space group and the direction of the modulation determine the
subcell generators ${\op G}^1,\ldots,{\op G}^B$ and the matrix $\altmat S$ 
and thus should be the same
for a whole class of materials. On the other hand, the displacements of atoms in the
asymmetric unit and the coordinate transformations $f^{I\kappa}$ 
are determined by the local potential and thus by the specific material under 
investigation.  This allows a separation of effects of crystal symmetry  from
effects characteristic for a special substance.\par 
The symmetry-adaptation (\ref{equation:WSV}) provides
a two-way relationship between the $B$ sets of SALMs and the
IRs of $\frak G({\vec q})$:
1) In the superposition (\ref{equation:deltaRn}), 
a symmetry mode $\delta{\altvec R}$ transforming according to a given IR requires 
the inclusion of a few distinct sets of SALMs (e.g.\ $I=1,\ldots,d_s$) 
and the exclusion of the others.
2) The included sets, on the other hand, may lead, with different spatial modulations 
of the respective relevant local variables $Q^{I1}_{\vec n},\ldots$, to
displacements transforming according to different (but in general not all) IRs.\par
In the simplest case -- one relevant set of SALMs ($d_s=1$) and one degree of freedom in
the asymmetric unit ($d_a=1$) -- expansion (\ref{equation:Wexpansion}) simplifies to
\begin{displaymath}
\delta{\altvec R}= \sum_{\vec n} \delta{\altvec R}^1_{\vec n}(Q^{11}_{\vec n},0,\ldots,0)
\end{displaymath}
and the number of variables per unit cell is reduced from $3K$ to $1$ ($Q^{11}_{\vec n}$).
\par
%
%
\section{Derivation of microscopic symmetry-adapted Hamiltonians and relation to other models}
The potential is expanded in a power series of the $Q^{I\kappa}_{\vec n}$. The
terms containing only variables with given $\vec n$ form
the local potential $\Phi^{\text{loc}}(Q^{11}_{\vec n},\ldots,Q^{B,3K'}_{\vec n})$ for
cell $\vec n$. The general Hamiltonian of the system is then
\begin{eqnarray*}
{\op H}&=&\sum_{\vec n} \left[ \sum_{I,\kappa} \left(P^{I\kappa}_{\vec n}\right)^2 + 
\Phi^{\text{loc}}(Q^{11}_{\vec n},\ldots,Q^{B,3K'}_{\vec n})\right]\\
&&+\Phi^{\text{int}}(\ldots,Q^{I\kappa}_{\vec n},\ldots,Q^{J\lambda}_{\vec m},\ldots),
\end{eqnarray*}
where $P^{I\kappa}_{\vec n}$ is the momentum conjugate to $Q^{I\kappa}_{\vec n}$.
Averaging out all irrelevant generalized coordinates by taking the
trace over variables $I\not=1,\kappa\not=1$ yields an effective 
Hamiltonian
\begin{eqnarray}
\label{equation:Hamil1}
{\overline{\op H}}&=& {\rm tr}(\rho {\op H}) = \sum_{\vec n} \left[\left(P^{11}_{\vec n}\right)^2 + {\overline{\Phi}}^{\text{loc}}(Q^{11}_{\vec n})\right] \\
&&+{\overline{\Phi}}^{\text{int}}(\ldots,Q^{11}_{\vec n},\ldots,Q^{11}_{\vec m},\ldots).\nonumber
\end{eqnarray}
\par
In the case of two relevant sets of SALMs per cell ($d_s=2,\,I=1,2$), 
one gets the two-variable version
\begin{eqnarray*}
{\overline{\op H}}&=& \sum_{\vec n} \left[\left(P^{11}_{\vec n}\right)^2 + \left(P^{21}_{\vec n}\right)^2 + {\overline{\Phi}}^{\text{loc}}(Q^{11}_{\vec n},Q^{21}_{\vec n})\right] \\
&&\qquad\qquad+ {\overline{\Phi}}^{\text{int}}(\ldots,Q^{11}_{\vec n},Q^{21}_{\vec n},\ldots,Q^{11}_{\vec m},Q^{21}_{\vec m},\ldots).
\end{eqnarray*}
A form with the superscript $21$ replaced by $12$ is obtained in the case 
of one relevant set of SALMs per cell ($d_s=1,\,I=1$) but two relevant generalized 
coordinates ($d_a=2$) per asymmetric subcell.\par
The total number of relevant variables per cell is $d_a\cdot d_s$. The local potential
${\overline{\Phi}}^{\text{loc}}$ and the 
interaction potential ${\overline{\Phi}}^{\text{int}}$ 
are effective potentials. 
Due to the thermal averaging over the irrelevant variables, they depend, by their very 
definition, directly on temperature 
and stresses (which both determine the strains). Usually
${\overline{\Phi}}^{\text{loc}}$ and ${\overline{\Phi}}^{\text{int}}$ are
expanded up to forth and second order respectively and only interactions between 
first, second and third nearest neighbors are kept.\par
The Hamiltonian (\ref{equation:Hamil1}) models only the 
symmetry-breaking contribution per cell. The totally symmetric IR
of ${\frak G}_0$, which also contributes to the displacements, is always present but 
has been absorbed into the atom 
positions of new reference states, which can be viewed (for the whole range of $T$ and
$p$) as (unstable) normal phase configurations.\par
For modulated crystals with normal phase space group ${\frak G}_0=Pnma$ exhibiting a
pseudoperiodicity along the direction of modulation, half cells instead of 
crystallographic unit cells may be chosen.
The Hamiltonian (\ref{equation:Hamil1}) corresponds to a version of the DIFFOUR model 
with one degree of freedom per lattice site, or to the ANNNI model.
The case where two local modes (per half cell)
with different transformation behavior ($d_s=2$) are
incorporated corresponds to Chen and Walker's model or to the DIS model. Assuming,
in addition, two relevant ($d_a=2$)
generalized variables $Q^{I1}_{\vec n}$ and $Q^{I2}_{\vec n}$ 
for each of the two relevant sets of SALMs, one arrives at a model
corresponding to a version of the DIFFOUR model with four degrees of freedom per lattice
site. If $Q^{I1}_{\vec n}$ and $Q^{I2}_{\vec n}$ are assumed to describe only
displacements of the betaine and \chemical{CaCl_2} groups respectively, the
model discussed in refs.\cite{Hli96,Qui96} for BCCD is obtained.\par
%
\section{Introduction of the pseudo-spin formalism}
%
Space group operations can be considered to transform the generalized variables 
$Q^p_{\vec n}$ instead of the modes ${\altvec W}^p_{\vec n}$ [$p=(I\kappa)$]. 
The local potential may have 
equivalent minima: if there is a minimum at a certain set of
the variables $Q^p_{\vec n}$ then
every space group operation which transforms a cell into itself 
but changes the configuration $Q^p_{\vec n}$ 
generates another equivalent minimum.
A further simplification may then be introduced by projecting the
remaining continuous generalized local variables 
\begin{equation}
\label{equation:PSDefinition}
Q^p_{\vec n}= |Q^p_{\vec n}| \sign Q^p_{\vec n} =: |Q^p_{\vec n}| \sigma^p_{\vec n}.
\end{equation}
onto two-valued pseudo spin variables $\sigma^p_{\vec n}$ representing their signs.
Because they assume only the values $+1$ or $-1$, they are called pseudo spins. 
They should not be confused with usual spins nor are they localized at
a definite position in the cell.   A switch of $\sigma^p_{\vec n}$ 
from $+1$ to $-1$ or vice versa corresponds to a transition from one well to another and is 
connected with a collective motion of all particles in the unit according to the 
corresponding SALM ${\altvec W}^p_{\vec n}$. The motion in a single well as described
by $|Q^p_{\vec n}|$ occurs on a much shorter timescale and can therefore be averaged out.
The interactions (expanded up to second order in $Q^p_{\vec n}$)
are replaced by the averages 
over single wells (taken separately for every $|Q^p_{\vec n}|$):
\begin{displaymath}
\overline \Phi^{pq}_{{\vec n}{\vec m}} \cdot
\overline{|Q^p_{\vec n}|} \cdot
\overline{|Q^q_{\vec m}|} \cdot
\sigma^p_{\vec n} \sigma^q_{\vec m}
=: J^{pq}_{{\vec n}{\vec m}} \sigma^p_{\vec n} \sigma^q_{\vec m},
\end{displaymath}
with couplings $J^{pq}_{{\vec n}{\vec m}}= 
\overline \Phi^{pq}_{{\vec n}{\vec m}} \cdot \overline{|Q^p_{\vec n}|} \cdot 
\overline{|Q^q_{\vec m}|}$.
Since the latter are defined as thermal averages, they 
depend inherently on external parameters.
The procedure yields the following pseudo spin Hamiltonian:
\begin{equation}
\label{equation:PSHamiltonian}
{\op H}= \frac{1}{2} \sum_{{\vec n}{\vec m}} \sum_{pq} J^{pq}_{{\vec n}{\vec m}}
\sigma^p_{\vec n} \sigma^q_{\vec m}.
\end{equation}
Special systems discussed along these lines were the ANNNI model 
(one relevant mode per unit) or
the DIS model (two variables)\cite{Neu94a,Neu96}.
Symmetry considerations yield conditions for the dependence of the potential on the 
generalized variables, which have for example been taken into account
when formulating the AANNDI\cite{Kur86} or the DIS\cite{Ple94} model.\par
%

%
%

\section{Symmetry of the structural modulation}
Directly
below the transition temperature from the unmodulated high temperature phase to the 
modulated phases a sinusoidal structural modulation is formed, which 
can be described by a Bloch mode of distinct symmetry (symmetry mode). 
Upon further cooling, it becomes more and more rectangular. Thus 
higher harmonics or further 
symmetry modes need to be considered. In good approximation, the relative displacements 
in one unit (cell or half cell) do not change whereas the amplitudes 
vary from unit to unit. This enters in the present approach by using restricted 
(e.g.\ one mode) 
local displacements (SALMs) and modeling the sinusoidal modulation and
deviations from it by a variation of the amplitudes of the SALMs.\par
The local displacements are determined by the local potentials; hence
the relevant SALMs and the respective coupling parameters can be determined for example 
by ab initio-type calculations. 
With the methods of statistical mechanics, the resulting Hamiltonians yield 
the stable profiles of the model variables. The superpositions of 
SALMs described by these profiles correspond to a mixture of a few symmetry modes. Just 
below the normal phase, where the profile
is nearly sinusoidal, the primary contribution to the structural deformation of the 
crystal transforms according to an IR of $\frak G({\vec q})$.\par
%
%
\section{Conclusions}
Symmetry-adapted local modes defined for crystallographic (sub)cells 
represent a natural lattice theoretical basis for the formulation of model Hamiltonians for 
describing modulated systems. 
Symmetry considerations in terms of the group of the modulation 
vector lead to the selection of admissible local mode combinations. By well defined thermal
averaging, the lattice theoretical systems can be projected onto
corresponding pseudo spin models. The coupling constants of the latter depend,
by their very definition, on temperature and strains. Explicit
calculations along these lines would allow, for not too complicated systems, to go
all the way from  an atomistic theory to explicit statements 
on phase diagrams, polarizations etc.  In such a treatment
approximations are of course necessary and should be made (in the original atomistic theory 
as well as in the consecutive model calculations); but one should avoid ad hoc 
assumptions (fits) as far as possible.
\par




\begin{thebibliography}{10}

\bibitem{San90b}
D.G. Sannikov.
\newblock {\em Sov. Phys. JETP} {\bf 70}, 1144 (1990).

\bibitem{Rib90}
J.L. Ribeiro, J.C. Tol\'edano, M.R. Chaves, A.~Almeida, H.E. M\"user,
  J.~Albers, and A.~Kl\"opperpieper.
\newblock {\em Phys.\ Rev.\ B} {\bf 41}, 2343 (1990).

\bibitem{Che91a}
Z.Y. Chen and M.B. Walker.
\newblock {\em Phys.\ Rev.\ B} {\bf 43}, 5634 (1991).

\bibitem{Che91b}
Z.Y. Chen and M.B. Walker.
\newblock {\em Phys.\ Rev.\ B} {\bf 43}, 760 (1991).

\bibitem{Hli96}
J.~Hlinka, O.~Hernandez, M.~Quilichini, and R.~Currat.
\newblock {\em Ferroelectrics} {\bf 185}, 221 (1996).

\bibitem{Jan81a}
T.~Janssen and J.A. Tjon.
\newblock {\em Phys. Rev. B} {\bf 24}, 2245 (1981).

\bibitem{Ell61}
R.J. Elliott.
\newblock {\em Phys.\ Rev.} {\bf 124}, 346 (1961).

\bibitem{Sel88}
W. Selke.
\newblock {\em Physics Reports} {\bf 170}, 213 (1988).

\bibitem{Kur86}
M.~Kurzy{\'n}ski and M. Halawa.
\newblock {\em Phys.\ Rev.\ B} {\bf 34}, 4846 (1986).

\bibitem{Ple94}
M.~Pleimling and R.~Siems.
\newblock {\em Ferroelectrics} {\bf 151}, 69 (1994).

\bibitem{Ple97a}
M.~Pleimling, B.~Neubert, and R.~Siems.
\newblock {\em Z.\ Phys.\ B}, in print.

\bibitem{Lu90}
H.M. Lu and J.R. Hardy.
\newblock {\em Phys.\ Rev.\ B} {\bf 42}, 8339 (1990).

\bibitem{Etx92a}
I.~Etxebarria, J.M. Perez-Mato, and G.~Madariaga.
\newblock {\em Phys.\ Rev.\ B} {\bf 46}, 2764 (1992).

\bibitem{Sch93}
G.~Schaack.
\newblock {\em Acta Physica Polonica A} {\bf 83}, 451 (1993).

\bibitem{Qui96}
M.~Quilichini and J.~Hlinka.
\newblock {\em Ferroelectrics} {\bf 183}, 215 (1996).

\bibitem{Neu94a}
B.~Neubert, M.~Pleimling, T. Tentrup, and R.~Siems.
\newblock {\em Ferroelectrics} {\bf 155}, 359 (1994).

\bibitem{Neu96}
B.~Neubert and R.~Siems.
\newblock {\em Ferroelectrics} {\bf 185}, 95 (1996).

\end{thebibliography}
\end{document}